\documentclass[12pt]{article}
\usepackage{amsmath}
\usepackage{amsthm}
\usepackage{amssymb}
\usepackage{epstopdf}

\newtheorem{example}{Example}

\begin{document}
\title{\bf Adjusted Jackknife Empirical Likelihood} \vspace{.1mm}
\author{Ying-Ju Chen$^{1}$ and Wei Ning$^{2}$ \footnote{Corresponding author. Email:
wning@bgsu.edu} \\ $^1$ Department of Information Systems and Analytics\\ Miami University, Oxford, OH 45056, USA\\$^2$ Department of
Mathematics and Statistics\\Bowling Green State University, Bowling
Green, OH 43403, USA}
\date{}
\maketitle

\noindent Abstract: Jackknife empirical likelihood (JEL) is an effective modified version of empirical likelihood method (EL). Through the construction of the jackknife pseudo-values, JEL overcomes the computational difficulty of EL method when its constraints are nonlinear while maintaining the same asymptotic results for one sample and two-sample U statistics. In this paper, we propose an adjusted version of JEL to guarantee that the adjusted jackknife empirical likelihood (AJEL) statistic is well-defined for all the values of the parameter, instead of restricting on the convex hull of the estimation equation. The properties of JEL have been preserved for AJEL. \\


\section{Introduction}
Empirical likelihood has been extensively studied and applied to different fields since it was first introduced by Owen (1988,1990). It combines the
advantage of the flexibility from nonparametric part and the efficiency from parametric part with the establishment of Wilks' theorem. See Owen (2001) for details. To obtain the statistic based on the empirical likelihood, it involves the maximization of the nonparametric likelihood through the calculation of Lagrange multiplier subject to given constraints. However, when the constraints are either linear or can be switched to be linear, the maximization process of EL method is easy to be accomplished. When the constraints are nonlinear, it will face some computational challenges as Jing et al. (2009) pointed out. They used U-statistics as examples to illustrate the computational difficulty of EL method. To overcome this difficulty, Jing et al. (2009) proposed a method called jackknife empirical likelihood method (JEL). The key idea of JEL is to transfer the statistic of interest to a sample mean situation under the traditional EL through constructing jackknife pseudo-values, that is, transfer nonlinear constraints to linear constraints and then apply the traditional EL method. The most attractive property of JEL is its simplicity. Jing et al. also established Wilks' results of JEL statistic for one sample and two-sample U-statistics. Since then, the JEL method has been applied to revisit various statistical problems including the ones which have been studied by the EL method by constructing U-statistics for the parameter of interest. For example, Gong et al. (2010) and Yang and Zhao (2013) proposed the methods based on JEL to study the ROC curve with the construction of the confidence intervals. Yang et al. (2015) developed a procedure based on the JEL to make statistical inference of the regression parameters for censored data. Chen et al. (2015) established a testing procedure based on JEL for the equality of two mean residual life functions. Wang and Zhao (2016) considered the comparison of two Gini indices for independent data and paired data and established Wilks' theorem for both situations.

However, as Chen et al. (2008) pointed out, computing profile empirical likelihood function which
involves constrained maximization requires that the convex hull of the estimating equation must have the zero vector as an interior point. When the required computational problem has no solution, Owen (2001) suggested assigning~$-\infty$~to the log-EL statistic. Chen et al. (2008) mentioned there are two drawbacks by doing so. To remedy the drawback of EL method, Chen et al. (2008) proposed an adjusted empirical likelihood which adds an artificial term to guarantee the zero vector to be within the convex hull, therefore, the solution always exists. In their work, they have showed the asymptotic results of the AEL is the same as that of the EL. Meanwhile, this method can achieve the improved coverage probabilities without taking Bartlett-correction or bootstrap calibration.

In this paper, we adapt Chen's idea and propose an adjusted jackknife empirical likelihood (AJEL) which combines the advantage of AEL and JEL. We also establish the Wilks' theorem for one sample and two-sample U-statistics of AJEL. Since the proposed method works for one-sample and two-samples U-statistics in general, therefore, it can be applied for any procedure based on the JEL through U-statistics. The paper is organized as follows. The main results of AJEL will be introduced in section 2 with the proofs provided in Appendix. Simulations have been conducted in section 3 to compare the performance of AJEL and JEL under different scenarios. The results indicate that AJEL improves the coverage probabilities, especially for small sample sizes. We apply AJEL to two data sets to construct confidence intervals for parameters and compare them with the ones obtained from JEL. Some discussion and our future work are provided in section 5.

\section{Adjusted Jackknife Empirical Likelihood}
Let $Z_1, Z_2, \ldots, Z_n$ be a random sample from a d.f. F. A \emph{U}-statistic of degree $m$ with a symmetric kernel $h$ is defined by

\begin{equation*}
U_n =\left(
\begin{array}{c}
n\\
m
\end{array}
\right)^{-1} \sum\limits_{1\leq i_i<\cdots<i_m\leq n} h(Z_{i_1}, Z_{i_2},\ldots, Z_{i_n})
\end{equation*}

and $\theta=\mathbb{E}h(Z_{i_1}, Z_{i_2},\ldots, Z_{i_n})$ is the parameter of interest. The jackknife pesudo-values of $U_n$ are given by
\begin{equation*}
\widehat{V}_i=nU_n-(n-1)U_{n-1}^{(-i)}
\end{equation*}
where $U_{n-1}^{(-i)}:=U_{n-1}(Z_1,\ldots, Z_{i-1}, Z_{k+1}, \ldots, Z_n)$. One can check that $\mathbb{E}\widehat{V}_i=\theta$ and $U_n=\frac{1}{n}\sum\limits_{i=1}^n\widehat{V}_i$. Shi (1984) showed that $\widehat{V}_i$'s are asymptotically independent under mild conditions. Let $p=(p_1,p_2,\ldots, p_n)$ be a probability vector which assigns
probability $p_k$ to $\widehat{V}_k$. Based on Jing et al. (2009), the jackknife empirical likelihood, evaluated at $\theta$ is
\begin{equation*}
L(\theta)=\max \left\{
\prod_{i=1}^n p_i:\sum\limits_{i=1}^n p_i=1, \sum_{i=1}^n p_i(\widehat{V}_i-\theta)=0
\right\}
\end{equation*}
and the jackknife empirical log-likelihood ratio at $\theta$ is
\begin{equation*}
\ln R(\theta)=-\sum\limits_{i=1}^n \ln \left\{ 1+\lambda (\widehat{V}_i-\theta)\right\}
\end{equation*}
where $\lambda$ satisfies
\begin{equation*}
\frac{1}{n}\sum_{i=1}^n \frac{\widehat{V}_i-\theta}{1+\lambda(\widehat{V}_i-\theta)}=0.
\end{equation*}
Let $g(x)=\mathbb{E}(x, Z_2, \ldots, Z_m)-\theta$ and $\sigma_g^2=var(g(Z_1))$. Jing et al. have showed
\begin{equation*}
-2\ln R(\theta)\rightarrow \chi_1^2 \mbox{ in distribution.}
\end{equation*}
when $\mathbb{E}h^2(Z_1,Z_2, \ldots, Z_n)<\infty$ and $\sigma_g^2>0$.

Now we apply the AEL method to $\widehat{V}_i, i=1, 2, \ldots, n$. Define a pseudo value by
\begin{equation}
\widehat{V}_{n+1}=-a_nU_n=-\frac{a_n}{n}\sum_{i=1}^{n}\widehat{V}_i
\end{equation}
for some $a_n>0$ which is suggested to take~$a_n=\log n/2$~by Chen et al. (2008). However, they pointed out that as long as~$a_n=o_p(n^{2/3}),$~the first
order asymptotic properties of the original~$\log EL$~statistic will be preserved for AEL. The adjusted jackknife empirical (AJEL) likelihood ratio of $\theta$ is given by
\begin{equation}
R(\theta)=\sup \left\{\sum_{i=1}^{n+1} \ln (n+1)p_i: \sum_{i=1}^{n+1} p_i=1, \sum_{i=1}^{n+1} p_i(\widehat{V}_i-\theta)=0 \right\}.
\end{equation}

With the same conditions given by Jing et al. (2009), we can establish Wilks' theorem of $-2\ln R(\theta)$ as~$\chi^2_1.$~The brief proof is provided in Appendix. Similarly, for two-sample U-statistics considered by Jing et al. (2009)
\begin{align*}
U_{n_1,n_2} &=\binom{n_1}{m_1}^{-1} \binom{n_2}{m_2}^{-1} \\
            & \times \sum_{1\le i_1<\cdots <i_{m_1}\le n_1} \sum_{1\le j_1<\cdots <j_{m_2}\le n_2}   h(X_{i_1},\cdots,X_{i_{m_1}},Y_{j_1},\cdots,Y_{j_{m_2}})\\
            &=U_n(X_1,\cdots, X_{n_1}, Y_1,\cdots, Y_{n_2}),
\end{align*}
we define
\begin{align*}
\widehat{V}_{n+1}=-a_nU_n=-\frac{a_n}{n}\sum_{i=1}^{n}\widehat{V}_i,
\end{align*}
where~$\widehat{V}_i=nU_n-(n-1)U_{n-1}^{(-i)},$~~$n=n_1+n_2$~and~$a_n=o_p(n^{2/3})>0.$~Then the AJEL likelihood ratio of~$\theta$~is defined as
\begin{align*}
R(\theta)=\sup \left\{\sum_{i=1}^{n+1} \ln (n+1)p_i: \sum_{i=1}^{n+1} p_i=1, \sum_{i=1}^{n+1} p_i(\widehat{V}_i-E\widehat{V}_i)=0 \right\}.
\end{align*}
With the same conditions in Jing et al. (2009), we can establish the Wilks' theorem for$-2\ln R(\theta)$ as~$\chi^2_1.$~

\section{Simulation Study}
In this section, we conduct a simulation study to illustrate the performance of the proposed method. Monte Carlo simulations with 1000 repetitions are performed to calculate coverage rate and average length of confidence intervals. To make fair comparisons with the JEL developed by Jing et al. (2009), we choose the same distributions as Jing et al. (2009) used under various scenarios. We use~$a_n=\log(n)/2$~in the definition of~$\widehat{V}_{n+1}.$~

\begin{table}
\caption{Probability weighted moment $E(XF(X)): X\sim \chi^2_1, \theta=0.8183$}
\label{weight}
\begin{tabular}{lcrrrr}\\ \hline \hline
\multicolumn{2}{l}{} & \multicolumn{2}{c}{\underline{$1-\alpha=0.90$}}  & \multicolumn{2}{c}{\underline{$1-\alpha=0.95$}}\\
\multicolumn{2}{l}{Nominal level} & Coverage(\%) & Length ($10^{-2}$) & Coverage (\%) & Length ($10^{-2}$)\\ \hline
$n=20$ & JEL & 83.8 (1.17) & 80.47  & 88.4 (1.01) & 94.09 \\
      & AJEL & 86.3 (1.09) & 88.02  & 90.7 (0.92) & 103.71\\
$n=30$ & JEL & 84.4 (1.15) & 67.15 & 89.6 (0.97) & 80.20\\
      & AJEL & 87.0 (1.06) & 71.60 & 91.3 (0.89) & 85.79 \\
$n=50$ & JEL & 86.0 (1.10) & 53.54 & 91.4 (0.89) & 65.00\\
      & AJEL & 87.8 (1.03) & 55.86 & 93.1 (0.80) & 67.90\\ \hline
\end{tabular}
\end{table}

\begin{example}\label{example1}(Probability Weighted Moment) Let $X_1, X_2, \ldots, X_n$ be a random sample from $\chi_1^2$ distribution and we denote~$F$~to be this distribution function. Then the probability weighted moment is $\theta=\mathbb{E}[X_1F(X_1)]=0.8183$ and the sample probability weighted moment is a \emph{U}-statistic with the kernel function $h(x,y)=\max \left\{x,y\right\}/2$.
\end{example}
Table \ref{weight} shows the coverage probabilities and average lengths of confidence intervals based on AJEL and JEL approaches for the Example \ref{example1}. As we can see in the table when the sample size is relatively small,~$n=20$~or~$n=30,$~AJEL improves the coverage probabilities by around 2.5\% comparing to those of JEL with slightly larger of average lengths of confidence intervals. When the sample size is moderate large,~$n=50,$~the performances of AJEL and JEL are similar. Overall, the AJEL surpasses the JEL in the coverage probability with the slightly longer average length.

\begin{example}\label{example2} (Receiver Operating Characteristic Curve) Let $X_1, X_2, \ldots, X_{n1}$ be independently and identically distributed to $Exp(1)$ and $Y_1, Y_2, \ldots, Y_{n_2}$ be independently and identically distributed to $Exp(1/9)$, respectively. Then the area under the receiver operating characteristic curve for diagnostic tests or biomarkers with continuous outcome is $\theta=P(Y_1>X_1)=0.9$ and a consistent estimator of $\theta$ based on the \emph{U}-statistic is given by
\begin{equation*}
U_n=\frac{1}{n_1n_2}\sum\limits_{i=1}^{n_1}\sum_{j=1}^{n_2}I\left\{Y_j>X_i\right\}.
\end{equation*}
\end{example}
The results in Table \ref{ROC} show that the coverage probability of AJEL is about 1\% higher than that of JEL when the sample size is $(n_1, n_2) = (10,10)$ or $(20,20)$ and two approaches are comparable in terms of coverage probability when $(n_1, n_2)=(35, 30)$. In conclusion, the AJEL outperforms the JEL in terms of the coverage probability with slightly longer average length.

\begin{table}
\caption{ROC curve $P(X<Y): X\sim Exp(1), Y\sim Exp(1/9), \theta=0.9$}
\label{ROC}
\begin{tabular}{lcrrrr}\\ \hline \hline
\multicolumn{2}{l}{\underline{Nominal level}} & \multicolumn{2}{c}{\underline{$1-\alpha=0.90$}}  & \multicolumn{2}{c}{\underline{$1-\alpha=0.95$}}\\
$(n_1, n_2)$ & & Coverage(\%) & Length ($10^{-2}$) & Coverage (\%) & Length ($10^{-2}$)\\ \hline
$(10, 10)$    & JEL & 81.4 (1.23) & 22.58  & 83.7 (1.17) & 27.43 \\
             & AJEL & 82.3 (1.21) & 24.70 & 83.9 (1.16) & 30.20 \\
$(15, 15)$    & JEL & 83.5 (1.17) & 19.00 & 87.8 (1.03) & 23.43 \\
             & AJEL & 86.0 (1.10) & 20.26 & 88.5 (1.01) & 25.06 \\
$(35, 30)$    & JEL & 92.3 (0.84) & 15.18 & 95.0 (0.69) & 18.06 \\
             & AJEL & 92.9 (0.81) & 15.70 & 95.2 (0.68) & 18.70 \\ \hline
\end{tabular}
\end{table}

\section{Real Data Example}
Duchenne Muscular Dystrophy (DMD) is a genetically transmitted disease which causes muscle degeneration and weakness. It usually affects boys only since it comes from the loss of a piece of DMD gene on the X-chromosome. Since a female carrier of DMD has no apparent symptoms in general, the suspicion of DMD may come from an affected boy in the family. It is well-known that carriers of DMD tend to have high levels of some serum enzymes or proteins, such as creatine kinase (CK), hemopexin (H), lactate dehydroginase (LD) and pyruvate kinase (PK). Andrews and Herzberg (1985) reported the data which recorded the levels of CK, H, LD, and PK for 75 blood samples of carriers and 134 samples from noncarriers for studying the detection of Muscular Dystrophy carriers.

In this section, we apply the adjusted jackknife empirical likelihood approach to the data set reported from Table 38.1 of Andrews and Herzberg (1985) which is the same data set in Jing et al. (2009).

\subsection{Creatine Kinase Level}
Let $X$ be the CK levals in noncarriers of DMD and $Y$ be the CK levels in carriers of DMD. Then the parameter of interest is $\theta=P(X<Y)$ and the point estimate of $\theta$ is
\begin{equation*}
\widehat{\theta}=\frac{1}{134\times 75}\sum_{i=1}^{134}\sum_{j=1}^{75} I(X_i<Y_j)=0.8635821.
\end{equation*}

Based on the AJEL approach, the 90\% and 95\% confidence intervals (CIs) for $\theta$ are $(0.8101, 0.9071)$ and $(0.7984, 0.9145)$. While applying the JEL method, the corresponding 90\% and 95\% CIs are $(0.8108, 0.9065)$ and $(0.7992, 0.9139)$.

\subsection{Comparison of Creatine Kinase and Hemopexin levels}
One may be interested in knowing if the measurement of CK levels is more efficient than that of other levels or which measurement is more efficient than others. For simplicity and convenience, we compare the CK and H levels which is also the case Jing et al. (2009) studied. Now, let $X^1$ and $X^2$ be the CK and H levels in noncarriers of DMD, respectively, and $Y^1$ and $Y^2$ be the CK and H levels in carriers of DMD, respectively. Then $P(X^1<Y^1)$ and $P(X^2<Y^2)$ could help to study the diagnostic accuracy of CK and H levels, respectively. The parameter of interest is $\theta \equiv P(X^1<Y^1)-P(X^2<Y^2)$ and the corresponding point estimator is
\begin{equation*}
\widehat{\theta}=\frac{1}{134\times 75} \sum_{i=1}^{134}\sum_{j=1}^{75} \left(I(X_i^1-Y_j^1)-I(X_i^2-Y_j^2)\right) = 0.1942289.
\end{equation*}

The 90\% and 95\% confidence intervals (CIs) for $\theta_1-\theta_2$ are $(0.1065, 0.2813)$ and $(0.0890, 0.2983)$ based on the AJEL approach. While applying the JEL method, the corresponding 90\% and 95\% CIs are $(0.1076, 0.2801)$ and $(0.0904, 0.2969)$. Both results support that the measurement of CK levels is more efficient than that of H levels.

\section{Discussion}
The jackknife empirical likelihood (JEL) method proposed by Jing et al. (2009) overcomes the potential computational difficulty in empirical likelihood (EL) when the constraints are nonlinear, while preserving Wilks' theorem. The adjusted empirical likelihood (AEL) method proposed by Chen et al. (2008) guarantees the existence of the solution of the required maximization of likelihood function with the establishment of Wilk's theorem. In this paper, we combine the advantage of these two methods to develop the adjusted jackknife empirical likelihood (AJEL). We establish Wilks' theorem for the proposed AJEL method for one sample and two-sample U-statistics. Simulations indicates that our method improves the coverage probability comparing to JEL method, especially when the sample sizes are small. Such a method is applied to two real data to construct the confidence intervals for the parameters. Since the Wilks' theorem has been derived for the AJEL on one-sample and two-samples U-statistics, therefore, the proposed method can
be applied to any other procedures based on the JEL through U-statistics and improve the accuracy of the coverage probability obtained by these procedures.

We only discuss the AJEL method based on U-statistics and develop corresponding properties. In our future work, we are interested in extending the similar idea to more general class of statistics than U-statistics and investigate the properties of AJEL method.

\section*{Acknowledgements}
The authors would like to thank Professor Jiahua Chen for kindly letting us use their programming codes as reference.

\section*{Appendix}
The proof of Wilk's theorem of the AJEL for one-sample U-statistics is similar to the proof of the AJEL for two-samples U-statistics. Therefore we only provide the proof of the former case. \\

\noindent Define $g_i=\widehat{V}_i-\theta$ where $\widehat{V}_i=nU_n-(n-1)U_{n-1}^{(-i)}$ which is given by Jing et al. (2009).
Let $g^*=\max\limits_{1\leq i \leq n} |\widehat{V}_i-\theta|$. Then $g^*=o_p(n^{1/2})$ through Lemma A.4 by Jing et al. (2009).\\
$\bar{g}_n=\frac{1}{n}\sum\limits_{i=1}^n g_i=\frac{1}{n}\sum\limits_{i=1}^n (\widehat{V}_i-\theta)=\frac{1}{n}\sum\limits_{i=1}^n \widehat{V}_i-\theta$.
Since $U_n=\frac{1}{n}\sum\limits_{i=1}^n\widehat{V}_i$, $\bar{g}_n=U_n-\theta$. Therefore, from Lemma A.2 (Jing et al. 2009) we obtain
\begin{equation*}
\frac{\sqrt{n}(U_n-\theta)}{2\sigma_g} \xrightarrow{d} N(0,1) \Rightarrow \bar{g}_n=O_p(n^{-1/2}).
\end{equation*}

As below, we show that $\lambda=O_p(n^{-1/2})$, where $\lambda$ is the solution of
\begin{equation*}
\sum\limits_{i=1}^{n+1}\frac{g_i}{1+\lambda g_i}=0.
\end{equation*}
Let $\rho=\|\lambda\|$ and $\widehat{\lambda}=\frac{\lambda}{\rho}$.
\begin{eqnarray*}
  0 &=& \frac{\widehat{\lambda}}{n}\sum\limits_{i=1}^{n+1} \frac{g_i}{1+\widehat{\lambda}g_i} \\
    &=& \frac{\widehat{\lambda}}{n}\sum\limits_{i=1}^{n+1} g_i-\frac{\rho}{n}\sum\limits_{i=1}^{n+1} \frac{(\widehat{\lambda}g_i)^2}{1+\rho\widehat{\lambda}g_i} \\
    &\leq& \widehat{\lambda}\bar{g}_n(1-\frac{1}{n}a_n)-\frac{\rho}{n(1+\rho g^*)}\sum\limits_{i=1}^n (\widehat{\lambda}g_i)^2.
\end{eqnarray*}

With $a_n=o_p(n)$, $g^*=o_p(n^{1/2})$, $\bar{g}_n=O_p(n^{-1/2})$, and $\frac{1}{n}\sum\limits_{i=1}^ng_i^2=\frac{1}{n}\sum\limits_{i=1}^n (\widehat{V}_i-\theta)^2=4\sigma_g^2+o(1)$ (lemma A.3 in Jing et al. 2009), we have
\begin{eqnarray*}
  \frac{\rho}{1+\rho g^*} &\leq& \widehat{\lambda}\bar{g}_n (1-\frac{1}{n}a_n)\cdot \left(\frac{1}{n}\sum\limits_{i=1}^n (\widehat{\lambda}g_i)^2\right)^{-1}\\
    &=& \widehat{\lambda}\bar{g}_n(1-\frac{1}{n}a_n)\left(\frac{1}{n}\sum\limits_{i=1}^n g_i^2 \right)
\end{eqnarray*}
since $\widehat{\lambda}=\frac{\lambda}{\rho}$ implies that $\|\widehat{\lambda}\|=1$.\\
That is,
\begin{eqnarray*}
  \frac{\rho}{1+\rho g^*} &\leq& \widehat{\lambda}'\bar{g}_n \left(1-\frac{a_n}{n}\right)\left(\frac{1}{n}\sum\limits_{i=1}^ng_i^2\right)^{-1} \\
    &=&  O_p(n^{-1/2})(1-o_p(1))(4\sigma_g^2+o_p(1))^{-1}\\
    &=& O_p(n^{-1/2}).
\end{eqnarray*}
Therefore, $\rho=\|\lambda\|=o_p(n^{1/2})$.
Denoting $\widehat{V}_n=\frac{1}{n}\sum\limits_{i=1}^n g_i^2=\frac{1}{n}\sum\limits_{i=1}^n (\widehat{V}_i-\theta)^2$, we have
\begin{eqnarray}
  0 &=& \frac{1}{n}\sum\limits_{i=1}^{n+1} \frac{g_i}{1+\lambda g_i}  \nonumber \\
    &=& \frac{1}{n}\sum\limits_{i=1}^{n+1} g_i-\frac{1}{n}\sum\limits_{i=1}^{n+1}\frac{\lambda g_i^2}{1+\lambda g_i} \nonumber \\
    &=& \bar{g}_n-\lambda \cdot \frac{1}{n} \sum\limits_{i=1}^{n+1}\frac{g_i^2}{1+\lambda g_i} \nonumber \\
    &=& \bar{g}_n-\lambda \cdot \frac{1}{n} \sum\limits_{i=1}^{n} g_i^2+o_p(n^{-1/2})\nonumber \\
    &=& \bar{g}_n-\lambda \widehat{V}_n + o_p(n^{-1/2}).  \label{equ3}
\end{eqnarray}
Therefore,
\begin{equation}
\lambda=\widehat{V}_n^{-1}\bar{g}_n+o_p(n^{-1/2}).
\end{equation}
We expand
\begin{eqnarray*}
-2\ln R(\theta_0;a_n)&=&2\sum\limits_{i=1}^{n+1} \ln (1+\lambda g_i)\\
 &=& 2\sum\limits_{i=1}^{n+1} (\lambda g_i-\frac{1}{2}(\lambda g_i)^2)+o_p(1).
\end{eqnarray*}
Replace $\lambda$ by $\widehat{V}_n^{-1}\bar{g}_n+o_p(n^{-1/2})$. Then
\begin{eqnarray*}
  -2\ln R(\theta_0;a_n) &=& n\bar{g}_n^2\widehat{V}_n^{-1}+o_p(1) \\
   &=& \frac{n(U_n-\theta)^2}{\frac{1}{n}\sum\limits_{i=1}^n (\widehat{V}_i-\theta)^2}+o_p(1)\\
   &\rightarrow& \chi_1^2.
\end{eqnarray*}
Here we give the detail of obtaining (\ref{equ3}).
\begin{eqnarray*}
\frac{1}{n}\sum\limits_{i=1}^{n+1}\frac{g_i}{1+\lambda g_i}  &=& \frac{1}{n}\sum\limits_{i=1}^{n+1}\left(g_i-\frac{\lambda g_i^2}{1+\lambda g_i}\right)\\
&=&\frac{1}{n}\left[\sum\limits_{i=1}^n g_i+g_{n+1}-\sum\limits_{i=1}^{n+1}\frac{\lambda g_i^2}{1+\lambda g_i}\right]\\
&=& \bar{g}_n+\frac{1}{n}g_{n+1}-\frac{1}{n}\sum\limits_{i=1}^{n+1}\frac{\lambda g_i^2}{1+\lambda g_i}\\
&=& \bar{g}_n+\frac{1}{n}g_{n+1}-\frac{\lambda}{n} \left[\sum\limits_{i=1}^n g_i^2-\sum\limits_{i=1}^n \frac{\lambda g_i^3}{1+\lambda g_i}+\frac{g_{n+1}^2}{1+\lambda g_{n+1}}\right]\\
&=& \bar{g}_n-\frac{1}{n}\lambda \sum\limits_{i=1}^n g_i^2+\frac{1}{n}g_{n+1}+\frac{\lambda}{n}\sum\limits_{i=1}^n \frac{\lambda g_i^3}{1+\lambda g_i}-\frac{\lambda}{n}\frac{g_{n+1}^2}{1+\lambda g_{n+1}}\\
&=& \bar{g}_n-\lambda \widehat{V}_n+\frac{1}{n}g_{n+1}+\frac{\lambda}{n}\sum\limits_{i=1}^n \frac{\lambda g_i^3}{1+\lambda g_i}-\frac{\lambda}{n}\frac{g_{n+1}^2}{1+\lambda g_{n+1}}
\end{eqnarray*}
First, by Chen et al. (2008) we have
$g_{n+1}=-a_n\bar{g}_n=o_p(n)\cdot O_p(n^{-1/2})=o_p(n^{1/2})$. Therefore, we have $\frac{g_{n+1}}{n}=o_p(n^{-1/2})$. Second, by the Lemma A.4 of Jing et al. (2009), $\frac{1}{n}\sum\limits_{i=1}^n g_i^3=\frac{1}{n}|\widehat{V}_i-\theta|^3=o(n^{1/2})$. Also with $\lambda=O_p(n^{-1/2})$, $g_i^*=\max |g_i|=o_p(n^{1/2})$, we have
\begin{equation*}
\frac{\lambda}{n}\sum\limits_{i=1}^n \frac{\lambda g_i^3}{1+\lambda g_i}=O_p(n^{-1})\cdot o_p(n^{1/2})=o_p(n^{-1/2}).
\end{equation*}
Finally, since $g_{n+1}=o_p(n^{1/2})$ implies $g_{n+1}^2=o_p(n)$,
\begin{equation*}
\frac{\lambda g_{n+1}^2}{n}=\frac{O_p(n^{-1/2})\cdot o_p(n)}{n}=o_p(n^{-1/2}).
\end{equation*}
Also, $\lambda g_{n+1} = O_p(n^{-1/2})\cdot o_p(n^{1/2})=o_p(1)$. Then
\begin{eqnarray*}
\frac{\lambda}{n}\frac{g_{n+1}^2}{1+\lambda g_{n+1}}&=&\frac{\lambda g_{n+1}^2}{n}\frac{1}{1+\lambda g_{n+1}}\\
   &=&o_p(n^{-1/2})\cdot O_p(1)=o_p(n^{-1/2}) .
\end{eqnarray*}

Therefore, we prove (\ref{equ3}).

\begin{thebibliography}{99}

\bibitem{} Chen, J., Variyath, A.M. and Abraham, B. (2008). Adjusted Empirical Likelihood And Its Properties. \textit{Journal of Computational and Graphical Statistics,} {\bf{17}}(2), 426-443.

\bibitem{} Chen, Y.J., Ning, W. and Gupta, A.K. (2015). Jackknife Empirical Likelihood Test for Equality of Two Mean Residual
Functions. \textit{Communications in Statistics-Theory and
Methods.} Accepted.

\bibitem{} Gong, Y., Peng, L. and Qi, Y. (2010). Smoothed Jackknife Empirical Likelihood Method for ROC Curve. \textit{Journal of Multivariate Analysis,}
{\bf{101},} 1520-1531.

\bibitem{} Jing, B.Y., Yuan, J.Q. and Zhou, W. (2009). Jackknife Empirical Likelihood. \textit{Journal of the American Statistical Association,} {\bf{104}}, 1224-1232.
\bibitem{} Owen, A.B. (1988). Empirical Likelihood Ratio Confidence Intervals for a Single Functional. \textit{Biometrika,} {\bf{75}}, 237-249.
\bibitem{} Owen, A.B. (1990). Empirical Likelihood Ratio Confidence Reigons. \textit{The Annals of Statistics,} {\bf{18}}, 90-120.
\bibitem{} Owen, A.B. (2001). \textit{Empirical Likelihood.} New York: Chapman \& Hall/CRC.

\bibitem{} Wang, D. and Zhao, Y. (2016). Jackknife Empirical Likelihood for Comparing Two Gini indices. \textit{The Canadian Journal of Statistics,}
{\bf{44}}(1), 102-119.

\bibitem{} Yang, H., Liu, S. and Zhao, Y. (2015). Jackknife Empirical Likelihood for Linear Transformation Models with Right Censoring. \textit{Annals of the Institute of Statistical Mathematics,} DOI: 10.1007/s10463-015-0528-7.

\bibitem{} Yang, H. and Zhao, Y. (2013). Jackknife Empirical Likelihood Confidence Intervals for the Difference of two ROC Curves. \textit{Journal of Multivariate Analysis,} {\bf{115},} 270-284.



\end{thebibliography}
\end{document}